\documentclass[twocolumn]{aastex62}

\newcommand{\myemail}{speacock@lpl.arizona.edu}

\usepackage{romanbar}

\graphicspath{{./}{figures/}}

\accepted{2018 December 14}

\submitjournal{ApJ}

\shorttitle{EUV-IR Spectrum of TRAPPIST-1}
\shortauthors{Peacock et al.}

\begin{document}

\title{PREDICTING THE EXTREME ULTRAVIOLET RADIATION ENVIRONMENT OF EXOPLANETS AROUND LOW-MASS STARS: THE TRAPPIST-1 SYSTEM}

\email{\myemail}

\author{Sarah Peacock}
\affil{University of Arizona, Lunar and Planetary Laboratory, 1629 E University Boulevard, Tucson, AZ 85721, USA}

\author{Travis Barman}
\affiliation{University of Arizona, Lunar and Planetary Laboratory, 1629 E University Boulevard, Tucson, AZ 85721, USA}

\author{Evgenya L. Shkolnik}
\affil{School of Earth and Space Exploration, Arizona State University, Tempe, AZ 85281, USA}

\author{Peter H. Hauschildt}
\affil{Hamburger Sternwarte, Gojenbergsweg 112, D-21029 Hamburg, Germany}

\author{E. Baron}
\affil{Homer L. Dodge Department of Physics and Astronomy, University of Oklahoma, 440 W. Brooks, Rm 100, Norman, OK 73019-2061 USA}
\affil{Hamburger Sternwarte, Gojenbergsweg 112, D-21029 Hamburg, Germany}

\begin{abstract}

The high energy radiation environment around M dwarf stars strongly impacts the characteristics of close-in exoplanet atmospheres, but these wavelengths are difficult to observe due to geocoronal and interstellar contamination. On account of these observational restrictions, a stellar atmosphere model may be used to compute the stellar extreme ultraviolet (EUV; 100 -- 912 \AA) spectrum. We present a case study of the ultra-cool M8 dwarf star, TRAPPIST-1, which hosts seven transiting short-period terrestrial sized planets whose atmospheres will be probed by the \textit{James Webb Space Telescope}. We construct semi-empirical non-LTE model spectra of TRAPPIST-1 that span EUV to infrared wavelengths (100 \AA\ -- 2.5 $\mu$m) using the atmosphere code PHOENIX. These upper-atmosphere models contain prescriptions for the chromosphere and transition region and include newly added partial frequency redistribution capabilities. In the absence of broadband UV spectral observations, we constrain our models using HST Ly$\alpha$ observations from TRAPPIST-1 and \textit{GALEX} FUV and NUV photometric detections from a set of old M8 stars ($>$1 Gyr). We find that calibrating the models using both data sets separately yield similar FUV and NUV fluxes, and EUV fluxes that range from (1.32 -- 17.4) $\times$ 10$^{-14}$ ergs s$^{-1}$ cm$^{-2}$. The results from these models demonstrate that the EUV emission is very sensitive to the temperature structure in the transition region. Our lower activity models predict EUV fluxes similar to previously published estimates derived from semi-empirical scaling relationships, while the highest activity model predicts EUV fluxes a factor of ten higher. Results from this study support the idea that the TRAPPIST-1 habitable zone planets likely do not have much liquid water on their surfaces due to the elevated levels of high energy radiation emitted by the host star.

\end{abstract}


\NewPageAfterKeywords

\section{Introduction} \label{sec:intro}
High energy radiation is damaging to close-in exoplanets, as increased levels of exposure can cause atmospheric expansion and escape and lead to the loss of global oceans \citep{Lammer2007,Owen2012,Luger2015,Chadney2016}. Planets become vulnerable to water loss as stellar far-ultraviolet (FUV; 1150 -- 1700 \AA) fluxes dissociate atmospheric H$_2$O, yielding atomic hydrogen susceptible to ionization by stellar extreme ultraviolet (EUV; 100 -- 912 \AA) radiation \citep{kasting1985,miguel2015,Bolmont2017}. As the combined X-ray and EUV (XUV; 5 -- 912 \AA) radiation heats and expands a planet's upper atmosphere, mass loss occurs via the hydrodynamic outflow of hydrogen or in the form of ion pickup by the stellar wind \citep{tian2008, murrayclay2009, rahmati2014,tripathi2015}.

Current capabilities allow for FUV and limited X-ray (5 -- 175 \AA) measurements, but observing stars other than the Sun across the EUV is impossible due to a lack of instruments operating in the necessary wavelength range. Additionally, EUV observations are hindered by contamination from Earth's geocoronal hydrogen and helium gas and optically thick interstellar hydrogen absorbing most of the spectrum between 400 -- 912 \AA \ \citep{Barstow2007}. Under such restrictions, efforts to quantify EUV radiation rely on semi-empirical models that extrapolate into the EUV from either X-ray or UV observations \citep[e.g.][]{Lecavelier2007, sanzforcada2011, fontenla2016} or empirical scaling relationships based on existing X-ray \citep{Chadney2015} or Lyman $\alpha$ (Ly$\alpha$; 1215.67 \AA) \citep{Linsky2014} observations. Ly$\alpha$ is the strongest emitting line in the FUV and likely drives photochemistry in the upper atmospheres of planets \citep{trainer2006}.

\cite{sanzforcada2011} produced synthetic XUV spectra of 82 late-F to mid-M planet hosts using emission measure distribution coronal models based on \textit{XMM-Newton}, \textit{Chandra}, and ROSAT X-ray observations. While emission features of highly ionized lines found at X-ray and EUV wavelengths form in the corona, the EUV continuum and many EUV and FUV emission lines form at cooler temperatures in the transition region and chromosphere. Since X-ray measurements only yield information about the hottest temperature layers of the star, relying solely on these observations to estimate the EUV and FUV spectrum results in under-predicted line fluxes \citep{france2016, Louden2017}.

\begin{figure*}[t]
\includegraphics[scale=1.2]{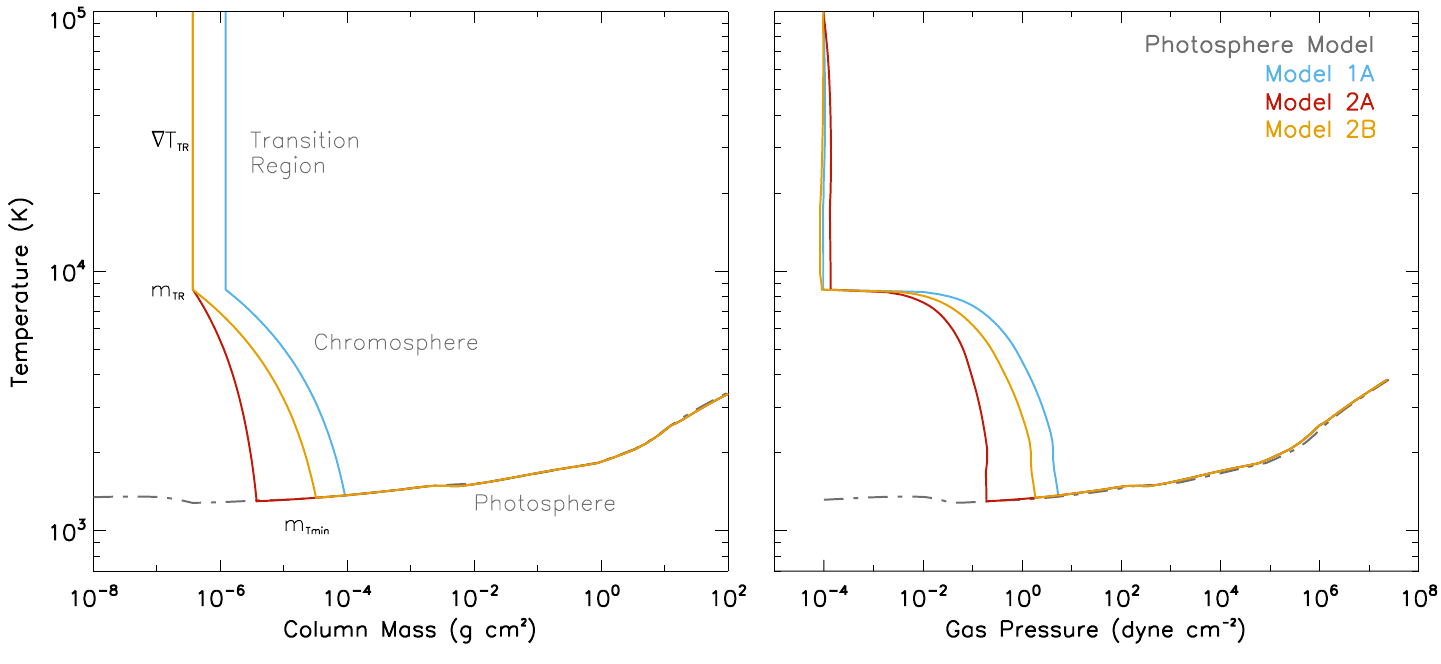}
\caption{Temperature structures corresponding to models with (solid lines) and without (dash-dot line) prescriptions for the chromosphere and transition region. Free parameters in the construction of the upper atmosphere are the column mass at the base and top of the chromosphere: \textit{m$_{Tmin}$} and  \textit{m$_{TR}$}, and the temperature gradient in the transition region: $\nabla$\textit{T$_{TR}$} (see labels in left panel for approximate locations). Models \textbf{1A} -- \textbf{2B} are described in Section \ref{sec:results} and their parameter values are given in Table \ref{tab:modelparam}.}
\label{fig:Tstructure}
\end{figure*}

Adapting their semi-empirical solar atmosphere code to apply to M stars, \cite{fontenla2016} used the Solar-Stellar Radiation Physical Modeling (SSRPM) tools to compute a 1D non-local thermodynamic equilibrium (non-LTE) spectrum of the M2 star, GJ 832. This M dwarf atmosphere model includes a prescription for the chromosphere, transition region, and corona and has direct comparison to observed spectra in FUV, NUV, and optical wavelengths. In this model, FUV spectra from the \textit{Hubble Space Telescope} (HST) were used to constrain the transition region structure, while the total X-ray luminosity and the formation temperature of the observed \ion{Fe}{12} 1242 \AA \ line were used to determine the temperature structure of the corona. The SSRPM model spectrum fits the observed UV and optical emission lines and continua well and predicts EUV luminosities similar to the active Sun.

The compact system around TRAPPIST-1 presents an interesting case study for examining the effects of high energy radiation on several close-in exoplanets. TRAPPIST-1 is a moderately active \citep{Gillon2016} ultracool dwarf star located at 12.1 pc, which hosts seven transiting planets orbiting within 6.3 $\times$ 10$^{-2}$ AU. Of these terrestrial-sized planets, three have surface equilibrium temperatures that could allow water to exist in liquid form on their surface (e, 0.0282 AU; f, 0.0371 AU; g, 0.0451 AU).

Recent studies estimate the XUV flux incident on the TRAPPIST-1 planets using X-ray and Ly$\alpha$ observations and analyze the stability of the planet atmospheres exposed to the predicted radiation levels. \cite{Wheatley2017} observed TRAPPIST-1 with \textit{XMM-Newton} and found that the X-ray luminosity of the star is similar to that of the quiet Sun. Using an $F_{EUV}/F_{X}$ scaling relationship from \cite{Chadney2015}, they estimate the XUV flux to be (6.8 -- 10.1) $\times$  10$^{-14}$ ergs s$^{-1}$ cm$^{-2}$. \cite{Bourrier2017a,Bourrier2017b} observed the Ly$\alpha$ line with HST in a series of four visits where the shape of the line profile appears to vary over time?. They calculate the EUV flux using the $F_{EUV}/F_{Ly\alpha}$ scaling relationship from \cite{Linsky2014} and summing with the observed X-
ray fluxes from \cite{Wheatley2017},  estimate the stellar XUV flux from TRAPPIST-1 to be (3.0 -- 4.1) $\times$  10$^{-14}$ ergs s$^{-1}$ cm$^{-2}$. Inputting these values into simple energy limited escape models, both studies find that there is sufficient high energy radiation to lead to the complete loss of oceans and atmospheres from the three habitable zone planets.

In this paper, we present new, non-LTE model EUV-IR spectra of the M8 star, TRAPPIST-1. The stellar EUV-NUV  spectrum is important for studying star-planet interactions and is a critical input for both photochemical and atmospheric escape models of exoplanet atmospheres. These models include prescriptions for the stellar upper atmosphere, including the chromosphere and transition region, where EUV, FUV and NUV fluxes originate. In Section \ref{sec:model}, we describe the construction of the model. We discuss how we constrain our models and compare resulting spectra to observations in Section \ref{sec:results}. In Section \ref{sec:discussion}, we explore the application of  the $F_{EUV}/F_{Ly\alpha}$ scaling relationship to our models and discuss current challenges in predicting high energy emission from low mass stars. Conclusions are given in Section \ref{sec:conclusions}.

\section{Model} \label{sec:model}

We construct 1D chromosphere models of TRAPPIST-1 using the atmosphere code PHOENIX \citep{hauschildt1993,hauschildt2006,baron2007}. This self-consistent multi-level non-LTE code is equipped with current atomic level data \citep{Dere1997,kurucz2014,delzanna2015,kurucz2017} suitable for the high temperatures and low densities found in M dwarf upper atmospheres and has been used in previous chromospheric investigations to model lines in the optical region of M dwarf spectra \citep{hauschildt1996,andretta1997, short1998,fuhrmeister2005, fuhrmeister2006}.

\begin{deluxetable}{cccc}[!bht]
\tablecaption{Model Parameters \label{tab:modelparam}}
\tablehead{
 \colhead{Model} &  \colhead{$\nabla$\textit{T$_{TR}$}} & \colhead{\textit{m$_{TR}$}} & \colhead{\textit{m$_{Tmin}$}}\\
 \colhead{} &  \colhead{(K g$^{-1}$ cm$^{-2}$)} & \colhead{(g cm$^2$)} & \colhead{(g cm$^2$)}
 }
\startdata
        \textbf{1A} & 10$^{9}$ & 10$^{-6}$ & 10$^{-4}$ \\
        \textbf{2A} & 10$^{8}$ & 10$^{-6.5}$ & 10$^{-5.5}$\\
        \textbf{2B} & 10$^{9}$ & 10$^{-6.5}$ & 10$^{-4.5}$\\
\enddata

\end{deluxetable}

Our PHOENIX models are computed in hydrostatic equilibrium on a log(column mass) grid (Figure \ref{fig:Tstructure}, Table \ref{tab:modelparam}). We begin with a photosphere-only model (gray, dash-dot line) in radiative-convective equilibrium that corresponds to the effective temperature, surface gravity, and mass of the star (Table \ref{tab:param}). The chemistry in the photosphere includes ions, molecules, and condensates, and is calculated directly using standard Gibbs free energy minimization. In this region, collisions dominate and LTE is an appropriate approximation for the atomic and molecular level populations. 

Increasing temperature distributions that simulate a chromosphere and transition region are superimposed to the underlying photosphere model, similar to the construction used in \cite{andretta1997}, \cite{short1998}, and \cite{fuhrmeister2005}. The chromosphere and transition region are characterized by high temperatures and low densities such that the collisional rates are low and radiative transfer is dominated by non-LTE effects.  In addition to including the LTE opacity from hundreds of millions of atomic and molecular transitions, PHOENIX is also capable of self-consistently modeling departures from LTE for many atoms and ions \citep{hauschildt1993,hauschildt2006,baron2007}.

\begin{deluxetable}{lc}[b!]
\tablewidth{100pt}
\tablecaption{Stellar Parameters \label{tab:param}}
\tablehead{}
\startdata
   		\textbf{Star} & \textbf{TRAPPIST-1}\\
        Spectral Type & M8 \\
         T$_{eff}$ (K) & 2559 $\pm$ 50\tablenotemark{1} \\
         M$_{\star}$ (M$_{\odot}$) & 0.080 $\pm$ 0.007\tablenotemark{1} \\
         R$_{\star}$  (R$_{\odot}$) & 0.117 $\pm$ 0.003\tablenotemark{1} \\
         Distance (pc) & 12.1 $\pm$ 0.4\tablenotemark{1} \\
         Age (Gyr) & 7.6 $\pm$ 2.2\tablenotemark{2}\\ 
\enddata
\tablerefs{(1) \citealt{Gillon2017} (2) \citealt{Burgasser2017}}
\end{deluxetable}

\begin{figure}[t]
\centering
\includegraphics[scale=0.6]{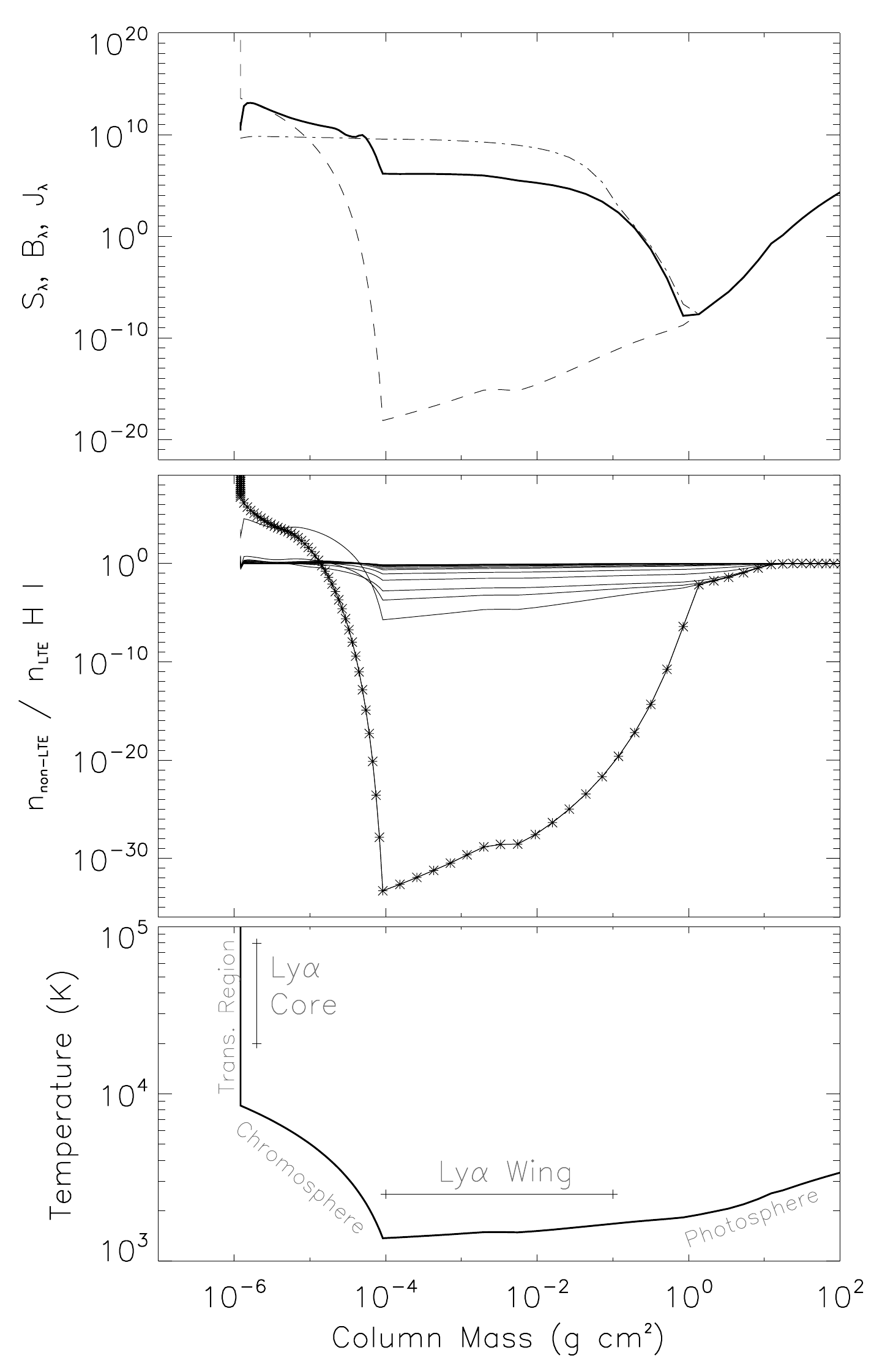}
\caption{\textit{Top Panel:} Radiative quantities for the center wavelength of Ly$\alpha$ in Model \textbf{1A}; Source function (S$_\lambda$, solid), Planck function (B$_\lambda$, dashed), mean intensity (J$_\lambda$, dash-dot). \textit{Middle Panel:} Departure coefficients for H I. Ground level indicated with asterisks. \textit{Bottom Panel:} Temperature structure for Model \textbf{1A} with line formation depths for Ly$\alpha$ wings and core indicated. \label{fig:lyman3}}
\end{figure}

For our models, we compute most light elements up to Ni in full non-LTE radiative transfer using species and background opacities provided by the PHOENIX and CHIANTI V8 \citep{delzanna2015} databases. These non-LTE calculations take into account 11,931 levels and 356,058 lines for 62 specifically considered ionization stages of the most abundant elements in the Sun: \ion{H}{1}; stages {\small\rmfamily I -- II\relax} of He, Ne, P; stages {\small\rmfamily I -- III\relax}  of Na, Mg, Al, S, Cl, Ca, Cr, Mn, Ni; stages {\small\rmfamily I -- IV\relax} of C, N, O, Si; and stages {\small\rmfamily I -- VI\relax} of Fe. 

The low density of the plasma in the upper atmosphere allows most emitted photons to escape with no further scattering such that pressure broadening is negligible in this region. This is not the case for strong resonance lines that form over extensive depths in a stellar atmosphere. In these lines, wings form in deeper atmospheric layers in LTE and are naturally broadened. The wings are controlled by the outward decrease of the source function, while the cores form in the upper atmosphere in non-LTE and only weakly depend on temperature and are instead more closely tied to the mean intensity (Figure \ref{fig:lyman3}). This can result in optically thick lines in emission with self-reversed cores (e.g. Ly$\alpha$, \ion{Mg}{2} \textit{h}\&\textit{k}, \ion{Ca}{2} H and K, H$\alpha$) \citep{Wood2005, fontenla2016,short1998,doyle1994}.

We employ a linear temperature rise with log(column mass) in our chromosphere and transition region that corresponds to a non-linear structure in pressure. This temperature-pressure profile is similar to the solar-inspired structure used by \cite{fontenla2016}, characterized by a steep lower-chromosphere followed by a temperature ``plateau" in the upper-chromosphere. If high resolution spectra are available, non-linear profiles can be tailored to fit individual lines, however, investigations have found that linear log(column mass) structures give better overall continuum fits \citep{fuhrmeister2005, andretta1997, eriksson1983}. Especially in the case of TRAPPIST-1, where the only UV spectral observation is the Ly$\alpha$ line \citep{Bourrier2017a, Bourrier2017b}, it is advantageous to utilize a linear temperature-log(column mass) structure in the chromosphere and transition region to predict the full UV spectrum.

\begin{deluxetable*}{lccccc}[!t]
\tablecaption{GALEX Photometry \label{tab:galex}}
\tablehead{
\colhead{} & \colhead{Spectral} & \colhead{Distance} & \colhead{T$_{eff}$} & \colhead{log(F$_{FUV}$)\tablenotemark{a}} & \colhead{log(F$_{NUV}$)\tablenotemark{a}} \\
\colhead{} & \colhead{Type} & \colhead{(pc)} & \colhead{(K)} & \colhead{(erg s$^{-1}$ cm$^{-2}$ \AA$^{-1}$)} & \colhead{(erg s$^{-1}$ cm$^{-2}$ \AA$^{-1}$)} 
}
\startdata
       M8 Field Stars&&&\\
        \hline
         2MASS 12590470-4336243 & M8 & 7.74 $\pm$ 0.07\tablenotemark{1} & 2570\tablenotemark{3} & $<$ -16.4 & -16.9 $\pm$ 0.2\\
        2MASS 10481463-3956062 & M8.5 & 4.05 $\pm$ 0.02\tablenotemark{1} & 2500\tablenotemark{4} & $<$ -17.0 & -17.1 $\pm$ 0.1\\
        2MASS 18353790+3259545 & M8.5 & 5.66 $\pm$ 0.02\tablenotemark{2} & 2578\tablenotemark{5}& $<$ -17.5 & -16.9 $\pm$ 0.01\\
        \hline
        HAZMAT III Sample\tablenotemark{6}\\
       \hline
        0.08-0.35 M$_\sun$, $\sim$ 5 Gyr & M3.5 -- M9 & $\cdots$ & $\cdots$ & -17.3$^{+0.4}_{-0.3}$ & -17.0$^{+0.3}_{-0.7}$\\
        \hline
      PHOENIX Models&&&\\
        \hline
       Model 1A & $\cdots$ & $\cdots$ & 2559 & -17.1 & -16.5 \\
       Model 2A & $\cdots$ & $\cdots$ & 2559 &  -17.1 & -17.2 \\
       Model 2B & $\cdots$ & $\cdots$ & 2559 &  -17.5 & -16.9 \\
\enddata
\tablecomments{F$_{FUV}$ and F$_{NUV}$ of the three M8 field stars and the HAZMAT III sample are from \textit{GALEX}. FUV values for the M8 field stars are upper limits. Uncertainties in the HAZMAT III sample represent inner quartiles of the full sample, which include detections and upper limits in both F$_{FUV}$ and F$_{NUV}$. Synthetic model FUV and NUV photometry are computed over the same wavelengths as the \textit{GALEX} filter profiles.}
\tablenotetext{a}{Flux density scaled to 12.1 pc}
\tablerefs{(1) \citealt{weinberger2016}; (2) \citealt{dupuy2012}; (3) \citealt{Pecaut2013}; (4) \citealt{Rajpurohit2013}; (5) \citealt{Rojas2012}; (6) \citealt{schneider2018}}

\end{deluxetable*}

We set the temperature at the top of the chromosphere to be between 8000 -- 8500 K since above these temperatures, hydrogen is fully ionized and is no longer an efficient cooling agent \citep{ayres1979}. This thermally unstable transition region extends steeply upwards until stability is reattained near the corona. In our models, we set the hottest layer to be 10$^5$ K, which corresponds to the top of the transition region.

Increasing the temperature in the outer layers through the addition of a chromosphere and transition region leads to large flux increases at shorter wavelengths. The chemistry switches from being dominated by molecules such as H$_2$ and CO, to atoms and ions in the lower chromosphere near 2000 K. The UV spectrum is characterized by strong emission lines of ionized species and bound-free edges from continuous opacity sources, particularly, the Lyman continuum and \ion{Si}{1} continuum. PHOENIX has been updated to include bound-free molecular opacities for H$_2$, CO, CH, NH, and OH, which are particularly important in shaping the NUV continuum in M stars \citep[and references within]{Fontenla2015}.

We adjust our model temperature structures through altering the location (\textit{m$_{Tmin}$}) and thickness of the chromosphere (\textit{m$_{TR}$}) and transition region ($\nabla$\textit{T$_{TR}$}). Higher UV emission is generated by attaching the chromosphere deeper in the atmosphere and decreasing the temperature gradient in the transition region. The integrated flux density across the FUV band (1340 -- 1811 \AA) is very sensitive to changes in all three parameters, while the NUV (1687 -- 3008 \AA)  is most sensitive to the depth at which the chromospheric temperature rise begins. Specifically, the entire UV pseudo-continuum\footnote{Pseudo-continuum is defined as the theoretical real continuum as affected by numerous molecular absorption features.} increases uniformly as the upper atmosphere is statically shifted towards higher column mass, while the FUV flux density changes by 2 -- 4 times that of the NUV when altering $\nabla$\textit{T$_{TR}$} or \textit{m$_{TR}$} alone. Increasing \textit{m$_{Tmin}$} leads to more overall UV flux, with flux densities increasing 2 -- 3 times more in the NUV band than in the FUV band.

\section{Model Comparisons}  \label{sec:results}

\begin{figure*}[htb!]
\includegraphics[scale=1.1]{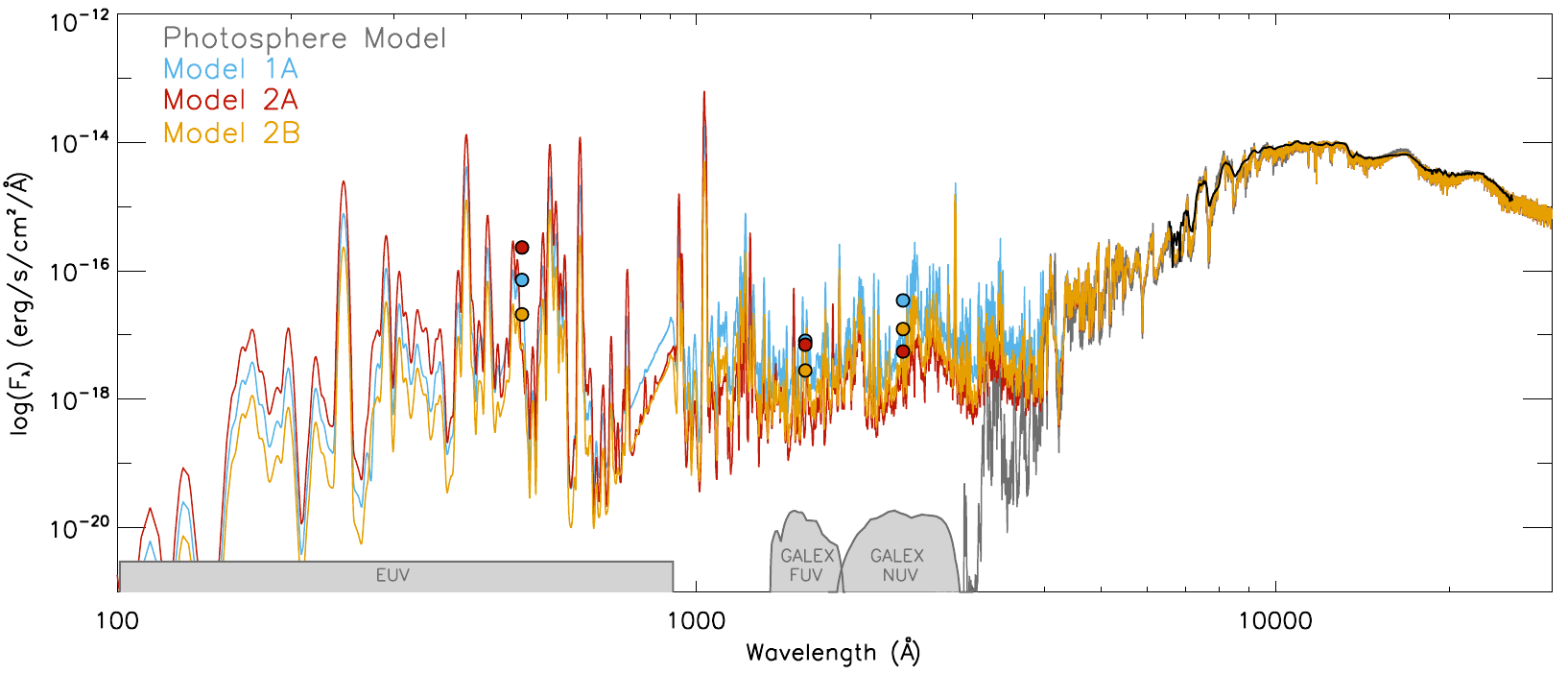}
\caption{Model EUV through IR spectra of TRAPPIST-1. Model \textbf{1A} (\textit{blue}) matches the TRAPPIST-1 Ly$\alpha$ reconstruction from \cite{Bourrier2017a}, Models \textbf{2A} and \textbf{2B} (\textit{2A: red, 2B: orange}) are calibrated to the range of \textit{GALEX} NUV detections of the M8 field stars and are consistent with the FUV upper limits. Spectral resolution in the models has been degraded for clarity. A model without a prescription for the chromosphere or transition region is plotted in gray; this photosphere-only model under-predicts the spectrum below 3000 \AA. Calculated EUV, FUV, and NUV synthetic photometry for the upper atmosphere models are plotted as circles. The filter profiles used to calculate these photometric points are shown in gray along the bottom axis. A SpeX NIR spectrum of TRAPPIST-1 \citep{Gillon2016} is plotted in black.\label{fig:spectrum}}
\end{figure*}

In the absence of broadband UV spectral or photometric observations of TRAPPIST-1, we use two methods to constrain the UV spectrum. First, we construct a model (\textbf{1A}) with the primary intent to replicate the reconstructed TRAPPIST-1 Ly$\alpha$ observation from \cite{Bourrier2017a}. For our second method, we use photometry from the \textit{Galaxy Evolution Explorer} (\textit{GALEX}) of the only three field-age M8 stars that have known distances and scale them to 12.1 pc (Table \ref{tab:galex}). We create a pair of models (\textbf{2A\&B}) to represent the range of NUV detections and the more rigorous FUV photometric upper limits, since all three stars are physically consistent with a log(F$_{FUV}$) $<$ -17.5.

 We also considered calibrating the models to the range of \textit{GALEX} FUV and NUV fluxes for a larger sample of field-age M stars (type M1 --M4) from the HAZMAT program \citep{Shkolnik2014}. However, in an extension of that work to lower-mass M stars (HAZMAT III), \cite{schneider2018} found that old ($\sim$ 5 Gyr) late-type M stars retain higher levels of UV flux compared to early M stars, indicating that the sample from \cite{Shkolnik2014} would not accurately represent UV emission from TRAPPIST-1. We instead examined the FUV and NUV flux densities for a large number of field-age late-M stars (0.08 -- 0.35 M$_\sun$) in HAZMAT III in order to refer to a more applicable sample of representative stars. The median values from these 56 FUV and 62 NUV detections from low-mass M3.5 -- M9 stars (Table \ref{tab:galex}) falls within the range of flux densities of the three M8 stars, so we proceeded with calibrating the models to the original sample.

Our PHOENIX spectra are presented in Figure \ref{fig:spectrum}; we compare the model that matches the Ly$\alpha$ observations (\textbf{1A}) to those that are calibrated to the \textit{GALEX} NUV detections from the three M8 stars (\textbf{2A\&B}). The corresponding model temperature structures are shown in Figure \ref{fig:Tstructure} with parameters listed in Table \ref{tab:modelparam}. A photosphere model computed with the T$_{eff}$, M$_{\star}$, and log(g) of TRAPPIST-1 is plotted as the dash-dot curve in Figure \ref{fig:Tstructure} and as the gray curve in Figure \ref{fig:spectrum}. We validate the stellar parameters used in this base photosphere model by comparing it to a near-infrared spectrum of TRAPPIST-1 \citep{Gillon2016}.  This photosphere-only model spectrum is dominated by molecular absorption features and vastly under-predicts the UV spectrum due to a lack of a prescription for the upper atmosphere.

At ultraviolet wavelengths, the optical depth reaches unity in the chromosphere where the source function deviates from the Planck function and non-LTE effects become important. Allowing for departures from LTE influences the emergent UV spectrum through increased flux levels in the Lyman continuum due to pumping effects of strong lines \citep{fuhrmeister2006}, but most notably through the decrease in the emergent flux of Ly$\alpha$ (top panel, Figure \ref{fig:lyman2}) and ionized chromospheric iron lines. Individual line profiles are impacted by the non-LTE treatment in both line strengths and the potential for self-reversal. These lines are typically narrower than LTE lines since they form higher in the atmosphere and are therefore not subject to pressure broadening. When treating our list of 62 species in non-LTE, the FUV flux density relative to a model computed fully in LTE decreases by a factor of 3, and the NUV flux density, which is dominated by contributions from iron lines, decreases by a factor of 50.

\subsection{Comparison to TRAPPIST-1 Ly$\alpha$ Observation}

\begin{figure}[!h]
\includegraphics[scale=0.55]{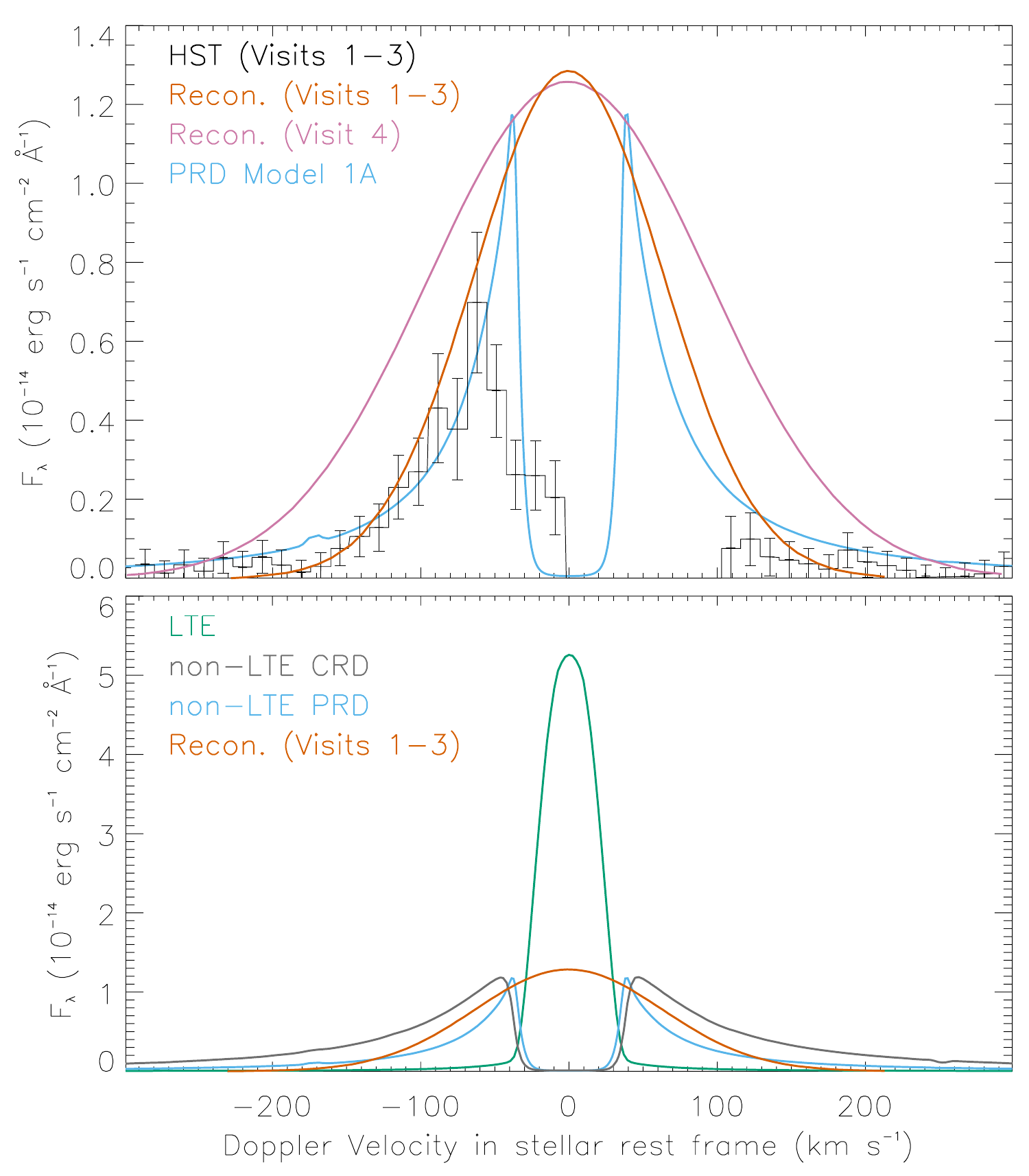}
\caption{ \textit{Top Panel:} Average of four raw HST observations of Ly$\alpha$ from Visits 1 -- 3 from \cite{Bourrier2017a} (black histogram). Overplotted: Line reconstructions from \cite{Bourrier2017a} for Visits 1 -- 3 (red) and from \cite{Bourrier2017b} for Visit 4 (pink) and Model \textbf{1A} non-LTE PRD profile (blue). \textit{Bottom Panel:} Computed Ly$\alpha$ profiles for Model \textbf{1A} calculated in LTE (green), non-LTE CRD (gray), and non-LTE PRD (blue). Line reconstruction from \cite{Bourrier2017a} for Visits 1 -- 3 overplotted in red. \label{fig:lyman2}}
\end{figure}

Our first method for constraining the UV spectrum for the model uses the only UV observations of TRAPPIST-1: reconstructed Ly$\alpha$ spectra from \cite{Bourrier2017a,Bourrier2017b}. The HST observations were taken with the G140M grating on STIS in four separate visits throughout 2016. The average of the raw spectra taken during Visits 1 -- 3 is plotted as the black histogram in the top panel of Figure \ref{fig:lyman2}. The Ly$\alpha$ wings did not vary significantly in the observations during these first visits, but differences in Visit 4 led to increased flux in the reconstructed wings, suggesting the shape of the line evolved throughout the four month observation period. \cite{Bourrier2017b} hypothesize that this variation could be caused by a trailing hydrogen exosphere from a transiting planet or could have stellar origins relating to the flow of hydrogen gas within the stellar atmosphere. Observations of Ly$\alpha$ for stars other than the Sun are contaminated by interstellar hydrogen and deuterium absorption near the line core and therefore only provide the intrinsic wing profiles. To correct for the geocoronal and ISM absorption, reconstructions must be performed on the observed line profiles \citep[e.g.,][]{Wood2005,france2013,Youngblood2016}. Single-component Gaussian reconstructions from \cite{Bourrier2017a,Bourrier2017b} of Visits 1-3 and Visit 4 are shown in the top panel of Figure \ref{fig:lyman2}.

Since hydrogen is the dominant atomic species in stellar atmospheres, line formation occurs at various depths and at lower densities than most other resonance lines. Non-LTE effects impacting both the size and shape of the Ly$\alpha$ line are seen in the bottom panel of Figure \ref{fig:lyman2}. At this wavelength scale, when treated in LTE, Ly$\alpha$ appears as a narrow line in full emission (green), vastly overpredicting the line center and underpredicting the wings of the reconstruction (red). When non-LTE calculations are considered (gray, blue), the line profile broadens and presents a self-reversed core. Emission lines with self-reversals occur when wings form deeper in the atmosphere, where the source function is increasing with temperature and the atmosphere is relatively close to thermal equilibrium, while the core is forming in the upper atmosphere, where departures from LTE are large and emerging photons are no longer coupled to the local temperature (Figure \ref{fig:lyman3}). 

When considering the model atmosphere calculations, in addition to non-LTE, partial frequency redistribution (PRD) becomes necessary for the accurate computation of radiative losses in strong resonance lines such as Ly$\alpha$. Complete frequency redistribution (CRD) accounts for overlapping radiative transitions, and is appropriate for computing most spectral lines. The addition of the PRD formalism becomes important in strong lines where coherent scattering is a major excitation mechanism (e.g., Ly$\alpha$, \ion{Mg}{2}, and \ion{Ca}{2}). As part of this work, PHOENIX is now equipped with PRD capabilities, implementing the methods detailed in \cite{uitenbroek2001} and \cite{hubeny1995}. Coherent scattering of photons largely affects the shape of the wings in the line profiles of strong resonance lines, as seen in the bottom panel of Figure \ref{fig:lyman2} where Ly$\alpha$ treated in CRD is plotted in gray versus in PRD in blue.

Although self-reversal in the Ly$\alpha$ emission line is seen in observations of both the quiet and active Sun \citep{Fontenla1988,Tian2009}, some reconstruction techniques (including those plotted in Figure \ref{fig:lyman2}) assume that the line is fully in emission and neglects the potential for a self-reversed core. To aid in the reconstruction of the Ly$\alpha$ line, it is possible to utilize observations of chromospheric \ion{Mg}{2} \textit{h} and \textit{k} (2794.5, 2802.3 \AA) emission lines. Similar to Ly$\alpha$, this NUV doublet is optically thick and the observations have to be corrected for interstellar contamination. While \ion{Mg}{2} \textit{h} and \textit{k} form at slightly lower temperatures than Ly$\alpha$, they have similar line profiles in the observed solar spectrum \citep{donnelly1994, lemaire1998}. \cite{Wood2005} reconstructs intrinsic Ly$\alpha$ emission lines for 33 cool stars by fitting the ISM absorption. They use \ion{Mg}{2} \textit{h} and \textit{k} lines to estimate the shape of the central portion of the Ly$\alpha$ profile, and the majority of the best fits to observed spectra for F -- K stars show a central reversal in both Ly$\alpha$ and \ion{Mg}{2} \textit{h} and \textit{k}, while those for M stars do not. Uncertainties in measurements, target variability, and stellar rotation period contribute to an overall uncertainty in F$_{Ly\alpha}$ of $\sim$30\% using this technique. Unlike the M stars in \cite{Wood2005}, high resolution HST observations of the M-dwarf GJ 832 show reversed cores in the \ion{Mg}{2} doublet, and along with the Ly$\alpha$ line, are modeled with a central reversal in the \cite{fontenla2016} GJ 832 M-dwarf upper atmosphere model. Obtaining similar resolution \ion{Mg}{2} \textit{h} and \textit{k} observations of TRAPPIST-1 would provide important information to help better estimate the central portion of the intrinsic  Ly$\alpha$ profile.

Another way to improve reconstructions of Ly$\alpha$ profiles and conclusively determine if M star Ly$\alpha$ cores contain inversions is through observations high radial velocity stars. Kapteyn's Star presents an ideal case where the M1 star has a high radial velocity ($\sim$ \texttt{+}245 km s$^{-1}$) such that the Ly$\alpha$ line is Doppler shifted 0.99 \AA \ away from the geocoronal emission feature. \cite{Guinan2016} observed the Ly$\alpha$ emission region (1214.5 -- 1217.5 \AA) with HST, and their ``stellar only" emission profile contains a very slight self-reversal. Conversely, \cite{Youngblood2016} fit this same observation with no self-reversal. Due to the ambiguity in these analyses, it would be beneficial to increase the number of high resolution Ly$\alpha$ observations of  high radial velocity M stars.

The Ly$\alpha$ line from Model \textbf{1A} is plotted in blue in Figure \ref{fig:lyman2} (top and bottom panels). The line profile matches the wings of the reconstruction for Visits 1 -- 3 well and displays a prominent self-reversal with the depth of the inverted core dropping to continuum flux levels, resulting from non-LTE effects. The depth of the central reversal is very sensitive to the thermal structure in the transition region in addition to other input physics. As seen in the top panel of Figure \ref{fig:lymaneuv}, Models \textbf{1A} and \textbf{2B} both have deep inverted cores and both have temperature gradients in the transition region of $10^9$ K g$^{-1}$ cm$^{-2}$, while Model \textbf{2A} has a much shallower reversal and a $\nabla$T$_{TR}$ of $10^8$ K g$^{-1}$ cm$^{-2}$. We suspect additional uncertainties in the line center may come from the lack of a corona in our model, which would photoionize the lower layers, affecting the collisional rates where the core is forming. The model-predicted Ly$\alpha$ integrated line flux is 38\% less than the reconstructed profile from Visits 1 -- 3 and 2.4 times less than that from Visit 4 (Table \ref{tab:Obs}).

Computing synthetic \textit{GALEX} photometry for Model \textbf{1A} yields an FUV flux density consistent with the M8 field stars and the inner quartiles of the HAZMAT III sample. The NUV flux density is slightly above these ranges, but within the total spread of the full HAZMAT III sample for old low mass M stars (Table \ref{tab:galex}).
\begin{deluxetable}{cccc}[t]
\tablewidth{\columnwidth}
\tabletypesize{\scriptsize}
\tablecaption{EUV and Ly$\alpha$ Fluxes \label{tab:Obs}}

\tablehead{
 \colhead{F$_{EUV}$} &  \colhead{F$_{Ly\alpha}$}&\colhead{Source}\\
 \colhead{(10$^{-14}$ erg s$^{-1}$ cm$^{-2}$)} &  \colhead{(10$^{-14}$ erg s$^{-1}$ cm$^{-2}$)}  & \colhead{}
 }
\startdata
        4.8 & 0.5 & Model 1A\\
       17.4 & 0.03 & Model 2A\\
       1.32 & 0.1 & Model 2B\\
       0.7$^{+0.3}_{-0.1}$ & 0.8$^{+0.3}_{-0.2}$&  \cite{Bourrier2017a}\\
       1.1 $\pm$ 0.1& 1.2 $\pm$ 0.1& \cite{Bourrier2017b}\\
        3.8 -- 5.8  & $\cdots$ &  \cite{Wheatley2017}\\
\enddata
\tablecomments{$\lambda_{EUV}$ = 100 -- 900 \AA, $\lambda_{Ly\alpha}$ = 1214.4 -- 1217 \AA. \\
F$_{Ly\alpha}$ for reconstructions from Visits 1 -- 3 and accompanying F$_{EUV}$ computed via the \cite{Linsky2014} scaling relationship are from \cite{Bourrier2017a}. F$_{Ly\alpha, EUV}$ for reconstructions from Visit 4 are from \cite{Bourrier2017b}. \cite{Wheatley2017} computes F$_{EUV}$ from an  $F_{EUV}/F_{X}$ scaling relationship and does not predict F$_{Ly\alpha}$. }

\end{deluxetable}

\subsection{Comparison to M8 Field Star UV Photometry}

In a second analysis, we construct models that match the NUV flux densities of the M8 field stars in Table \ref{tab:galex} and that are consistent with the more rigorous FUV upper limits. Model \textbf{2A} is tailored to the lower log(F$_{NUV}$) = -17.1 erg s$^{-1}$ cm$^{-2}$ \AA$^2$ and has slightly more FUV emission than Model \textbf{2B}, which has the higher log(F$_{NUV}$) = -16.9 erg s$^{-1}$ cm$^{-2}$ \AA$^2$ and an FUV flux density equal to the strictest upper limit. The resulting EUV flux densities span an order of magnitude, strongly tied to the factor of ten difference in their transition region temperature gradients.

\begin{figure*}[hbt!]
\includegraphics[scale=1.1]{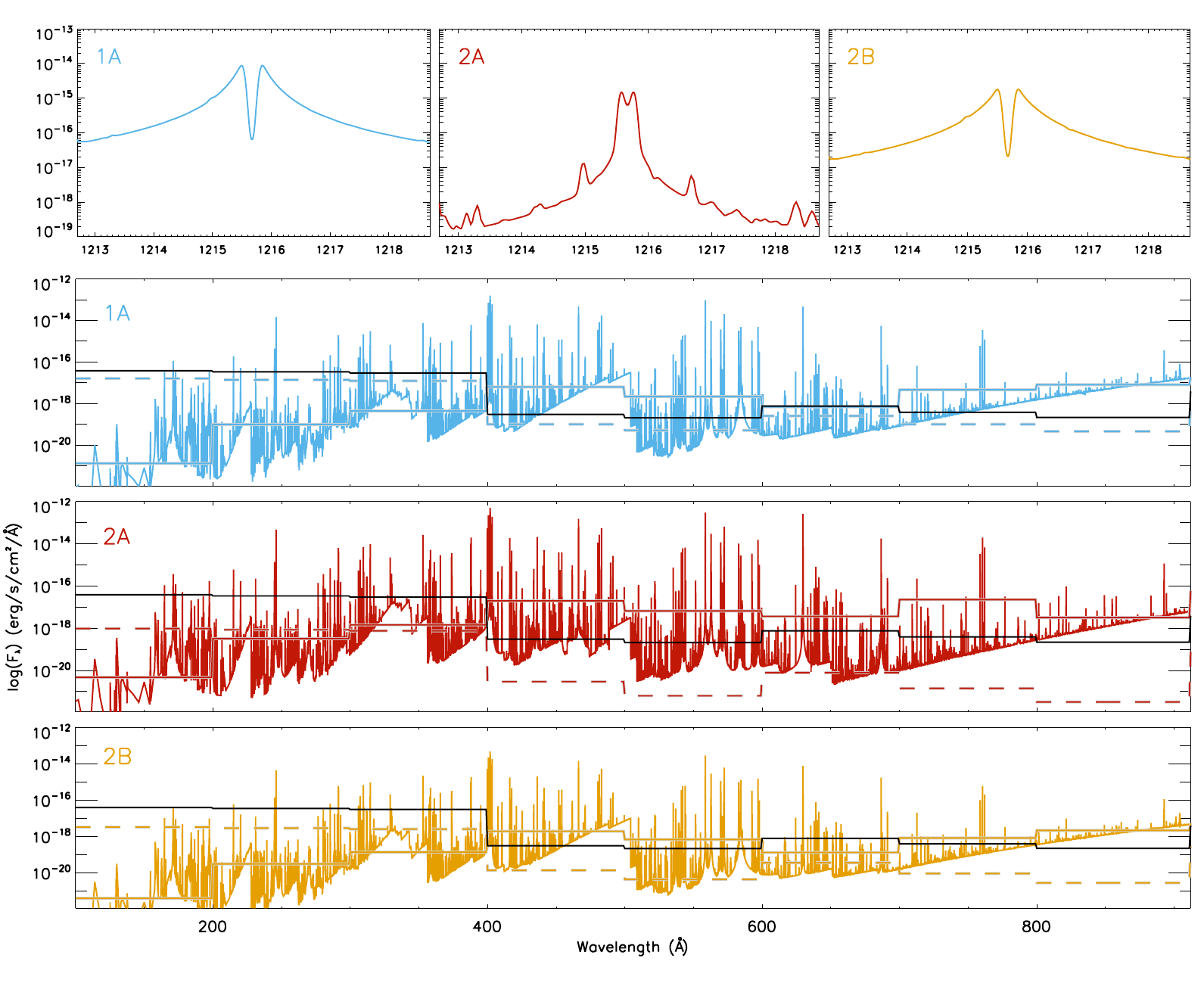}
\caption{ (\textit{top}) Model PRD Ly$\alpha$ profiles at the spectral resolution of STIS G140M.  (\textit{bottom}) Full resolution EUV spectra for Models \textbf{1A -- 2B}. Estimated EUV flux densities in 100 \AA \ wavelength bands using the \cite{Linsky2014} $F_{EUV}/F_{Ly\alpha}$ scaling relationship with F$_{Ly\alpha}$ from the Visits 1 -- 3 reconstruction from \cite{Bourrier2017a} overplotted in black. Average model spectra flux densities in 100 \AA \ wavelength bands overplotted in corresponding color (solid line). Estimated EUV flux densities using model F$_{Ly\alpha}$ overplotted in corresponding color (dashed line).}\label{fig:lymaneuv}
\end{figure*}

The EUV continuum and many strong emission lines at EUV and FUV wavelengths form in the transition region at temperatures between log(T) 4.3 -- 4.78 K \citep{Sim2005}. Examples include \ion{He}{1}  (584 \AA), \ion{C}{3} (977 \AA), \ion{H}{1} Ly$\beta$  (1025.7 \AA), \ion{C}{2} (1036.3, 1037 \AA), and \ion{Si}{4}  (1393.7, 1402.7 \AA). A decrease in $\nabla$\textit{T$_{TR}$} results in higher EUV continuum flux as well as more emergent line flux (e.g. \ion{O}{5} at 629.7, 760.2, 760.4, and 762.0 \AA). Full resolution EUV spectra of the models are shown in Figure \ref{fig:lymaneuv}.

The synthetic FUV photometry is calculated over the \textit{GALEX} FUV filter profile wavelength range and does not include Ly$\alpha$. The majority of the flux comes from a few strong emission lines, most notably: \ion{Al}{2}, \ion{Si}{2}, and \ion{Fe}{2}. The pseudo-continuum in the covered FUV wavelength range is shaped by bound-free opacities of Si, Mg, and Fe, while the NUV is shaped more by molecular opacity sources. In the NUV spectrum, most flux comes from the \ion{Mg}{2} \textit{h} and \textit{k} doublet, Al lines, and a forest of ionized Fe lines. Lines strengths for select lines that have formation temperatures in the chromosphere and transition are given in Table \ref{tab:lineflux}.

\section{Discussion} \label{sec:discussion}

The EUV fluxes from our upper atmosphere models range from (1.32 -- 17.4) $\times$ 10$^{-14}$ ergs s$^{-1}$ cm$^{-2}$ (Table \ref{tab:Obs}). Models \textbf{1A} and \textbf{2B} fall within the range of previous F$_{EUV}$ predictions derived from X-ray and Ly$\alpha$ observations. Model \textbf{2A}, which replicates the lower limit of the M8 field star NUV flux densities and has the lowest F$_{Ly\alpha}$, yields an EUV flux $\sim$ 10 times higher than previous upper estimates.

 The EUV continuum is dominated by bound-free edges from both \ion{H}{1} at 912 \AA \ and \ion{He}{1} at 504 \AA \ and contains many emission lines of highly ionized species. We caution that our EUV spectra are to be taken as upper limits for wavelengths $>$ 300 \AA. Many of the strong EUV emission lines form in the transition region, where radiative rates dominate and non-LTE effects are important. Our current models do not calculate every line in non-LTE, and as a result, the brightest lines in the model EUV spectrum: \ion{Fe}{7} at 246 \AA, \ion{Ne}{4} at 401.9 \AA, \ion{O}{5} at 629.7 \AA, and \ion{O}{6} at 1031.1 and 1037.6 \AA \ are narrow, but very strong LTE lines. Since there are no EUV observations of M8 stars to directly measure against, we estimate that compared to the  \cite{Bourrier2017a} predicted spectrum, in our most active model (\textbf{2A}),  the EUV spectrum for wavelengths $>$ 300 \AA \ could be overestimating the actual EUV flux by a factor of 20.

For wavelengths $<$ 300 \AA, the model spectra should be taken as lower limits. Our models do not include a corona, and therefore under-predict X-ray and some EUV continuum flux. Additionally, some highly ionized EUV emission lines that form in the corona, including Fe {\small\rmfamily X \& XI\relax}, \ion{Mg}{9}, and Ne {\small\rmfamily XIII \& X\relax} are calculated in LTE. Comparing the model spectra below 300 \AA \  to the predicted EUV flux from \cite{Wheatley2017}, we estimate that our lowest activity model, Model \textbf{2B}, could be under-predicting the flux in these wavelengths by up to a factor of 10$^3$.

\begin{deluxetable}{ccccc}[t]
\tablewidth{\columnwidth}
\tabletypesize{\scriptsize}
\tablecaption{Line fluxes (10$^{-18}$ erg s$^{-1}$ cm$^{-2}$) of select computed chromosphere and transition region lines \label{tab:lineflux}}

\tablehead{
 \colhead{Species} & \colhead{$\lambda$ (\AA)} & \colhead{Model 1A} & \colhead{Model 2A} & \colhead{Model 2B} 
 }
\startdata
	\ion{Si}{2} &	1264.73 & 110.83 &		0.12 &	24.45 \\
	\ion{Si}{3}	&	1298.95 & 0.29 &		0.05 &	0.10 \\
	\ion{O}{1}	&	1302.17 & 5.01 &		0.03 &	0.73 \\
	\ion{O}{1}	&	1304.86 & 4.52 &		0.02 &	0.48 \\
	\ion{O}{1}	&	1306.03 & 3.54 &		0.01 &	0.24 \\
	\ion{Si}{2} &	1309.28 & 103.09 &		7.91 &	26.08 \\
	\ion{C}{2}	&	1335.71 & 36.60 &		0.18 &	5.42 \\
	\ion{Fe}{2}	&	1391.08 & 210.72 &		36.87 &	75.94 \\
	\ion{Si}{2}	&	1526.71 & 44.29 &		0.09 &	5.92 \\
	\ion{Si}{2}	&	1533.43 & 47.67 &		0.20 &	6.94 \\
	\ion{C}{4}	&	1550.77 & 17.12 &		8.34 &	9.83 \\
	\ion{Al}{2}	&	1670.79 & 310.14 &		0.11 &	101.20 \\    
	\ion{Al}{1} & 1766.39 &   470.93 &		151.80 &	158.00 \\
	\ion{Mg}{2} \textit{k} &	2796.35 & 6919.41 &		1281.44 &	8274.69 \\
	\ion{Mg}{2} \textit{h} &	2803.53 & 11297.31 &	286.99 &	4432.94 \\
	\ion{Ca}{2} K &	3934.78 & 389.44 &		1.06 &	50.74 \\
	\ion{Ca}{2} H &	3969.59 & 302.74 &		0.79 &	34.15 \\
\enddata
\tablecomments{Computed line flux for Ly$\alpha$ is given in Table \ref{tab:Obs}.}
\end{deluxetable}

\subsection{F$_{EUV}$/F$_{Ly\alpha}$}

We compare our model EUV spectra to estimated EUV flux densities calculated in 100 \AA \ wavelength bands using the \cite{Linsky2014} F$_{EUV}$/F$_{Ly\alpha}$ empirical scaling relationship in Figure \ref{fig:lymaneuv}. The EUV spectrum using the Ly$\alpha$ reconstructed flux of Visits 1 -- 3 from \cite{Bourrier2017a} is plotted in black. The scaling relationship is derived from solar observations and is used for predicting F$_{EUV}$ for earlier type stars than TRAPPIST-1 (F5 -– M5), but generally agrees well with our model spectra, especially Models \textbf{1A} and \textbf{2B}. The largest discrepancies occur $<$ 300 \AA \ where the model spectra yield $\sim$ 10$^2$ -- 10$^5$ times less flux than the scaling relationship. Between 300 -- 900 \AA, the models yield on average 1 -- 20 times more flux than the scaling relationship, mostly stemming from continuum flux in the \ion{H}{1} bound-free edge (800 -- 900 \AA) and the overestimated LTE lines.

We calculate the line flux from our model Ly$\alpha$ over the same wavelength range as \cite{Bourrier2017a} and compute the 100 \AA \ EUV wavelength band fluxes (plotted as dashed lines in corresponding colors in Figure \ref{fig:lymaneuv}). While our Model \textbf{1A} Ly$\alpha$ matches the wings of the \cite{Bourrier2017a} reconstruction, the deep inverted cores in Models \textbf{1A} and \textbf{2B} and the more narrow line profiles of Model \textbf{2A} result in line fluxes that are 75 -- 85\% less than the Visits 1 -- 3 reconstruction. Calculating the broadband EUV spectrum with our model F$_{Ly\alpha}$ yields EUV fluxes that are a factor of 1.6 -- 35 lower than when using the reconstructed Ly$\alpha$ observation. Comparing the model EUV average flux density bins (solid colored lines) to their F$_{EUV}$/F$_{Ly\alpha}$ predicted bins, the discrepancies increase at wavelengths $>$400 \AA, especially for Model \textbf{2A}, where the scaling relationship predicts $\sim$ 10$^2$ -- 10$^5$ times less flux than the spectrum. This large difference in Model \textbf{2A} is likely due to the narrow line shape and small line flux in Ly$\alpha$.

\subsection{Challenges}

It is difficult to constrain precise models with a single data point. For example, Models \textbf{1A} and \textbf{2A} have identical FUV flux densities, but both their NUV and EUV fluxes differ by approximately one order of magnitude. Model \textbf{1A}, which is constrained by a single emission line, is potentially overpredicting the NUV, although these observations are of different stars and it is unknown if they occurred when the stars were in a similar activity state. Correct estimates of stellar EUV flux are important for studying the photochemistry and stability of exoplanet atmospheres, but  there are several challenges in attempting to derive these values: \\

\textit{Stellar Activity}: M stars are prone to flare in the ultraviolet \citep{fossi1996, hawley2003}, with the largest flares elevating the continuum emission by up to 200$\times$ quiescent levels \citep{loyd2018}. Late-type M stars remain UV active for much longer than their early-M counterparts, typically with more variability in the FUV than the NUV \citep{Miles2017}. This results in increased stellar variability even for field-age stars and can be seen in the evolving line profile in Visits 1 -- 3 versus Visit 4 of the TRAPPIST-1 Ly$\alpha$ observations. With the increased potential for UV activity in late-type M stars of all ages, in order to create a realistic panchromatic stellar spectrum, it is crucial that the observations be taken while the star is in the same activity state. UV observations are critically important for constraining stellar upper atmosphere models, and long-duration UV monitoring, like that to be done with the NASA-funded Star-Planet Activity Research CubeSat (SPARCS), is important for advancing our understanding of M star upper atmospheres in providing much needed variability and flaring data \citep{Shkolnik2018,Ardilla2018}. \\

\textit {Extensions of empirical scaling relationships to late-type M stars}: Both the F$_{EUV}$/F$_{X}$ and F$_{EUV}$/F$_{Ly\alpha}$ scaling relationships are based on solar observations and while they have been used to predict EUV fluxes for early to mid-M stars, they have not been validated in applications to late-M stars. Comparing the estimates to our models, we find that in some cases the empirical scaling relationships yield EUV fluxes consistent with the model spectra, however, our model with the lowest Ly$\alpha$ flux yielded and EUV flux 10 times higher than predicted by the scaling relationships. \\

\textit{Lack of corona in our model}: The models presented in this paper do not include a prescription for a corona. The $\sim$ 10$^6$ K coronal layers are a major source of X-ray fluxes as well as some EUV flux. While the majority of EUV radiation originates in the transition region, the models lack the flux contribution from highly ionized lines that form at coronal temperatures. Additionally, the hot corona irradiates downwards onto underlying layers in the stellar atmosphere and can alter the radiative rates. We anticipate that the addition of a corona to the model will increase the EUV flux mainly through changes in the continuum  below 300 \AA. In a future paper, Peacock et al. (in prep), the impact of the corona is explored in detail.

\section{Conclusions} \label{sec:conclusions}
Using two separate datasets, TRAPPIST-1 Ly$\alpha$ reconstructions and \textit{GALEX} UV photometry from old M8 field stars, we obtain model spectra with EUV fluxes that are closely aligned with previous estimates. The model EUV fluxes range from (1.32 -- 17.4) $\times$ 10$^{-14}$ ergs s$^{-1}$ cm$^{-2}$, as compared to the (0.7 -- 5.8) $\times$ 10$^{-14}$ ergs s$^{-1}$ cm$^{-2}$ derived from X-ray and Ly$\alpha$ empirical scaling relationships. The model EUV spectra and line centers in the Ly$\alpha$ profiles demonstrate sensitivity to the temperature structure in the transition region and are likely to increase with the addition of a corona to the model.

Analyses based on the previous estimates find the XUV flux to be high enough to erode both oceans and atmospheres on the TRAPPIST-1 habitable zone planets over several billion years. Applying the observed X-ray fluxes summed with our model EUV fluxes to an energy limited escape model would yield the same or greater mass loss rates, further suggesting these planets likely do not have much liquid water on their surfaces.

\acknowledgments

We would like to thank B. Fuhrmeister, I. Short and J. Aufdenberg for useful discussions. We also thank A. Schneider for providing GALEX photometry fluxes and D. Bardalez Gagliuffi for providing the TRAPPIST-1 SpeX spectrum. We gratefully acknowledge the helpful comments from the anonymous referee. This work was supported by NASA Headquarters under the NASA Earth and Space Science Fellowship Program - Grant NNX15AQ94H.  An allocation of computer time from the UA Research Computing High Performance Computing (HPC) at the University of Arizona is gratefully acknowledged. Resources supporting this work were also provided by the NASA High-End Computing (HEC) Program through the NASA Advanced Supercomputing (NAS) Division at Ames Research Center. This work was supported in part by DFG GrK 1351. A portion of the calculations presented here were performed at the H\"ochstleistungs Rechenzentrum Nord (HLRN), and at the National Energy Research Supercomputer Center (NERSC), which is supported by the Office of Science of the U.S.  Department of Energy under Contract No. DE-AC03-76SF00098. We thank all these institutions for a generous allocation of computer time. E.B. acknowledges support from NASA Grant NNX17AG24G. E.S. appreciates support from the NASA Habitable Worlds grant NNX16AB62G.

\software{PHOENIX \citep{hauschildt1993,hauschildt2006,baron2007}}




\end{document}